\documentclass[aps,twocolumn,showpacs,amsmath,amssymb,pra,superscriptaddress,floatfix,longbibliography]{revtex4-1}

\usepackage{amsmath}     
\usepackage{graphicx}    
\usepackage{hyperref} 
\usepackage{xcolor} 
\usepackage{physics} 
\usepackage{subcaption} 
\usepackage{ragged2e}
\usepackage[justification=justified, singlelinecheck=false]{caption}
\usepackage{bm}
\usepackage{booktabs}
\usepackage{soul}

\begin{document}
\title {Effects of finite trapping on the decay, recoil, and decoherence of dark states of quantum emitter arrays}
\author{M. Eltohfa}
 \email{meltohfa@purdue.edu}
 \affiliation{
 Department of Physics and Astronomy, Purdue University, West Lafayette, Indiana 47906 USA
}
\author{F. Robicheaux}%
 \email{robichf@purdue.edu}
\affiliation{
 Department of Physics and Astronomy, Purdue University, West Lafayette, Indiana 47906 USA
}
\affiliation{Purdue Quantum Science and Engineering Institute, Purdue
University, West Lafayette, Indiana 47907, USA}
\date{\today}

\begin{abstract}
    The collective interaction of electronic excitations with the electromagnetic field in atomic arrays can lead to reduced decay rates, forming subradiant states with applications in quantum information and memories. By including quantized vibrational excitations, we examine the effects of finite trap strength and light-mediated forces on highly subradiant singly-excited states for two, three, and many atoms in a 1D waveguide or free space. For waveguide-coupled and tightly trapped atoms, the recoil energy from photon emission can reach a vibrational quantum, even in the Lamb-Dicke regime. For weakly trapped atoms, the vibrational wavepackets are shifted or distorted due to induced forces and uneven decay. These effects lead to a time-dependent decay rate, extra vibrational energy transfer, and mixing of different electronic and vibrational states. The resulting entanglement entropy and infidelity can be mitigated by decreasing the induced forces or increasing trap strength. For quantum information storage, these findings suggest optimal array configurations in geometry and polarization. Our results provide insights for quantum memories and atom array experiments.
\end{abstract}

\maketitle

\section{Introduction}
\label{sec:introduction}

Achieving controlled coupling between light and matter is essential for coherent quantum control and quantum information processing. To enable this, one of the tools is to utilize the collective interaction between atoms by placing them close to each other \cite{guimond2019subradiant,jen2024photon, hammerer2010quantum, PhysRevLett.119.053901, PhysRevLett.117.243601, PhysRevA.97.023833, bettles2020quantum}. Compared to individual decay rates, collective emission rates can be significantly enhanced or suppressed \cite{dicke1954coherence,gross1982superradiance,gross1976observation,scully2009collective,pellegrino2014observation,prasad2000polarium,devoe1996observation}. In circular atomic arrays with subwavelength separations, for example, the decay rate of the excitation eigenstates decreases exponentially with the number of atoms \cite{asenjo2017exponential}, which is useful for the storage of photons \cite{PhysRevResearch.4.013110}.

Furthermore, the collective decay can be used to control the radiation pattern or the direction of the emitted photons \cite{plankensteiner2017cavity,grankin2018free,ruks2024negative, PhysRevA.108.030101}. For example, the radiation could be suppressed to all but one selected channel where the atoms radiate efficiently \cite{asenjo2017atom}. Another example is that the geometry of a 2D array of atoms can be optimized such that the array acts like a perfect mirror \cite{shahmoon2017cooperative,bettles2016enhanced,rui2020subradiant}. Additionally, the smaller decay rate of the subradiant or dark states results in the emitted light spectrum having a narrower linewidth. This is utilized for collective cooling of atoms to their motional ground state better than individual atom cooling \cite{rubies2024collectively}.

Most of the theoretical studies of collective decay assume that atoms are perfectly localized and fixed in space. While this assumption captures many of the collective effects mentioned above, it does not directly deal with motional aspects which are present in experiment. The effect of the motion is most evident in highly subradiant configurations, which take a long time to decay allowing motional distortion to build up signficantly. While subradiant states exist for different numbers of electronic excitations, this paper only deals with \emph{singly-excited} states.


Because the cooperative decay dynamics depends on the separation between the atoms, several detrimental properties could arise: 1) perfectly dark states are impossible; the excitation will necessarily have a minimum decay rate that arises from the spread in the position of the atoms \cite{rusconi2021exploiting,rubies2024collectively, PhysRevLett.103.123903, guimond2019subradiant, guimond2019subradiant}. 2) if the atoms are placed very close together, there could be cooperative forces that impart momentum and move the atoms from their initial positions \cite{suresh2022atom}. 3) the induced dipole forces can also introduce entanglement between the internal and motional degrees of freedom leading to unwanted decoherence for the internal degrees of freedom \cite{shahmoon2019collective,suresh2022atom, apolín2025recoilinducederrorscorrectionphotonmediated,hybrid2025quantizedmotion}. This, for example, will affect the fidelity of the long-time stored photons introduced in Ref. \cite{asenjo2017exponential,PhysRevResearch.4.013110}. 4) a long lifetime of a dark state is usually accompanied by large recoil after photon emission \cite{suresh2022atom, suresh2021photon}. 5) the imparted momentum during decay (due to the induced forces) leads to additional energy that can lead to undesired heating of a cold atom array. This heating would necessitate additional cooling steps each time the array is manipulated\cite{kikura2025tamingrecoileffectcavityassisted}.

In this work, we aim to address these issues. By including the vibrational aspects of the atoms, we provide quantitative results and scaling behavior of various effects of finite traps that build on and extend previous lines of research. 1) In addition to the $t=0$ lower bound on decay rates set by the spread of Gaussian motional states \cite{rusconi2021exploiting,rubies2024collectively, PhysRevLett.103.123903, guimond2019subradiant, guimond2019subradiant}, we examine a non-trivial \emph{later time, $t>0$, dynamics} of highly subradiant singly-excited states, where the motional wavefunction is shifted or distorted. This requires going beyond the Lamb-Dicke limit or expansion in Refs. \cite{hybrid2025quantizedmotion,rubies2024collectively} and beyond the sudden approximation used in Ref. \cite{suresh2021photon}. In particular, for the two atom case in Sec.\ref{sec:two_atoms_in_1d_waveguide}, we consider regimes where atomic motion is so significant that a second-order expansion in the vibrational coupling is not convergent and keeping higher order terms is necessary. 2) We extend the calculation in \cite{suresh2022atom, suresh2021photon} of the vibrational energy gained by the atoms \emph{before} and \emph{after} the emission of a photon in the case of \emph{spread-limited} subradiant states 3) We study the decoherence and infidelity of a dark state arising due to the induced forces and explore ways to minimize such undesriable effects.


\section{Methods}
\label{sec:model_and_underlying_assumptions}
In this section, we describe the theoretical framework used to study the collective decay of atoms. Only the single-excitation scenario is considered here. Moreover, we note that we do not have any driving fields such as a laser term. We assume the initial state has been prepared by some mechanism for $t<0$ in an electronically excited state at $t=0$. To account for the motion of the atoms and the internal states, we use a density matrix formalism expanded in the vibrational states for the atoms' motion and the internal states, similar to the formalism in Refs. \cite{suresh2022atom,PhysRevA.111.013711}. We assume each atom is trapped in a harmonic well which is identical for both the ground and excited states. This removes possible undesirable effects during the \emph{single atom} excitation and de-excitation processes (see Refs. \cite{karanikolaou2024near, rui2020subradiant}) and lets us focus on motional effects due to the \emph{collective interaction} between the atoms.

\subsection{System}
\label{sec:system}
We consider $N$ atoms of equal mass $m$. Each atom $i$ has two internal states, $\ket{g_i}$ and $\ket{e_i}$, with a transition frequency $\omega_0$ and a spontaneous decay rate $\gamma_0$. The operators $\sigma_{i}^{+}$ and $\sigma_{i}^{-}$ are the raising and lowering operators of the internal states of atom $i$: $\sigma_{i}^{+} = \ket{e_i}\bra{g_i}$ and $\sigma_{i}^{-} = \ket{g_i}\bra{e_i}$.

\begin{figure}
    \centering
    \includegraphics[width=0.5\textwidth]{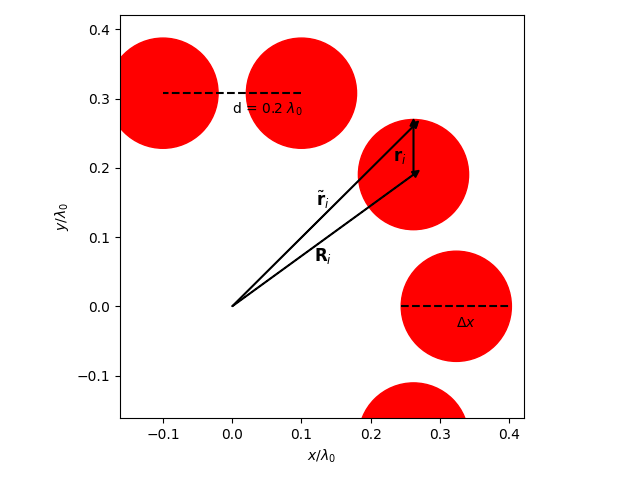}
    \caption{\justifying Schematic of part of an atom array on a circle. Here, the atoms are trapped regularly with separation $d = 0.2 \lambda_0$. The size of the red circle ($\Delta x$) indicates the spread (standard deviation) of the trap ground state. Position vectors to the center of the traps, $\textbf{R}_i$, and the absolute position of the atoms, $\tilde{\textbf{r}}_i$, as well as the relative position, $\textbf{r}_i$, are shown. }
    \label{fig:system}
\end{figure}

Each atom is trapped in a harmonic well. The location of the center of the trap of atom $i$ is denoted by $\textbf{R}_i$, while the location of the atom $i$ relative to its trap center is $\textbf{r}_i$. For the relative position of two traps, we use the notation $\textbf{R}_{ij} = \textbf{R}_i - \textbf{R}_j$ and the relative displacement as $\textbf{r}_{ij} = \textbf{r}_i - \textbf{r}_j$. We also use the absolute position of an atom which is denoted by $\tilde{\textbf{r}}_i = \textbf{R}_i + \textbf{r}_i$, with corresponding difference of two atoms $\tilde{\textbf{r}}_{ij} = \tilde{\textbf{r}}_i - \tilde{\textbf{r}}_j$. A schematic of the system is shown in Fig. \ref{fig:system}. For simplicity, the traps are of equal angular frequency $\omega_t$ and evenly separated by a distance $d = |\textbf{R}_{i+1} - \textbf{R}_{i}|$. The eigenstates of the trap for atom $i$ in each direction are denoted by $\ket{n_i}$ with energy $n_i\hbar \omega_t$ (where the conventional 1/2 is dropped for convenience). 

The density matrix of the system, $\rho$, can be expanded in the tensor product basis of the vibrational and internal states for each atom. We denote such a generic basis state by $\ket{A,V}$ where $A$ and $V$ are the collection of the internal and vibrational quantum numbers of the system, respectively: $\ket{A} =\ket{\alpha_1, \alpha_2, \dots}$, and $\ket{V}= \ket{n_1, n_2, \dots}$, where at most one $\alpha_j$ can be $e_j$. In this way, there are $N+1$ allowed internal states. Each $n_j$ can be $0,1,2,\dots$. Hence, the density matrix can be written as

\begin{equation}
    \rho = \sum_{A,V} \sum_{A',V'} \rho_{A,V,A',V'} \ket{A,V}\bra{A',V'}.
    \label{eq:density_matrix}
\end{equation}
 
For numerical feasibility, we can truncate the vibrational states to $0,1,2,\dots, n_{max}$, where we choose a phonon cut-off, $n_{max}$, to insure convergence of the numerical simulation. This way, we have $N_{vib} = n_{max} + 1$ vibrational states per atom. The vibrational Hilbert space will, therefore, contain $(N_{vib})^N$ states. The total number of states in the system is $N_{tot} = (N_{vib})^N (N+1)$.

\subsection{Master Equation}
\label{sec:equations_of_motion}

The evolution of the system's density matrix, $\rho$, is governed by a master equation
\begin{equation}
    \frac{d \rho(t)}{d t} = -\frac{\text{i}}{\hbar} \left[H_{\text{eff}}\rho(t)-\rho(t)H_{\text{eff}}^{\dagger}\right] + \mathcal{R}[\rho(t)].
    \label{eq:master_equation}
\end{equation} where $H_{\text{eff}}$ is an effective Hamiltonian that contains a trap term and a non-Hermitian interaction term, and $\mathcal{R}$ is a population recycling superoperator \cite{rubies2024collectively} defined below. The trap term is given by
\begin{equation}
    H_{\text{trap}} = \sum_{i=1}^N \hbar \omega_t a_i^{\dagger} a_i.
    \label{eq:trap_hamiltonian}
\end{equation} where $a_i$ is the lowering operator of the vibrational states of the trap of atom $i$.
The interaction term comes from the coupling with the electromagnetic field (which is traced out) and is given by 
\begin{equation}
    H_{\text{int}} = -\hbar\sum_{i,j} g(\tilde{\textbf{r}}_{ij}) \sigma_{i}^{+} \sigma_{j}^{-},
    \label{eq:interaction_hamiltonian}
\end{equation} where $g$ is a complex valued Green's function proportional to the electric field propagator and is given in  Eqs. (\ref{eq:1d_Greens_function}) and (\ref{eqn:Greens_3d}) below. The imaginary part of this Green's function gives a non-Hermitian interaction responsible for the collective decay of the excited state. On the other hand, the real part gives a Hermitian interaction responsible for the collective Lamb shifts of the energy of the states and also the induced forces which can shift the positions of the atoms before the decay of the excited state \cite{scully2009collective}.

The final term in Eq. (\ref{eq:master_equation}) is the recycling superoperator, $\mathcal{R}$, which accounts for the decayed population and the recoil energy during the photon emission step. The recycling superoperator acts on the density matrix in a relatively complicated way. The simplest description is in the position representation of the density matrix, where $\rho$ is expanded as
\begin{equation}
    \rho = \sum_{A,A'} \int dr dr' \rho_{A,A'}(r,r') \ket{A,r} \bra{A',r'}.
    \label{eq:density_matrix_position}
\end{equation} where $r$ and $r'$ are the collections of the positions of the atoms. The recycling superoperator is given by
\begin{equation}
    \begin{aligned}
    \mathcal{R}[\rho] = 2
    & \sum_{i,j} \sum_{A,A'} \int dr dr' \rho_{A,A'}(r,r') \,\text{Im}[g]\left(\tilde{\textbf{r}}'_{ij}\right) \\
    & \times \sigma_{i}^{-} \ket{A,r} \bra{A',r'}\sigma_{j}^{+},
    \end{aligned}
    \label{eq:recycling_superoperator}
\end{equation} where $\text{Im}[g]$ is the imaginary part of the Green's function, and $\tilde{\textbf{r}}'_{ij}=\tilde{\textbf{r}}_{i}-\tilde{\textbf{r}}'_{j}$. The multiplication of the density matrix by the position dependent Green's function in the above equation is responsible for and explains the mechanism of the imparting of momentum and energy to the atoms during the photon emission step. We use a spectral method employing the eigenstates of a truncated position operator $(a + a^{\dagger})$ as the basis to calculate this superoperator action. This results in a faster convergence with the number of basis states as done in the appendix of Ref. \cite{PhysRevA.111.013711}.

\subsection{Green's function}
\label{sec:Greens_function}
The Green's function in Eq. (\ref{eq:interaction_hamiltonian}) depends on the underlying geometry of the system. For atoms coupled to a 1D waveguide (assuming no other decay channel), the Green's function has a simple form and leads to coupling given by \cite{asenjo2017atom}:
\begin{equation}
    g(\textbf{r}) = \text{i}\frac{\gamma_0}{2}\exp\left(\text{i}k_0 |\textbf{r}|\right),
    \label{eq:1d_Greens_function}
\end{equation} where $k_0 = 2\pi/\lambda_0 = \omega_0/c$ is the wave number of light, $\lambda_0$ is the wavelength of light, and $c$ is the speed of light. For a waveguide along the x-axis, $\textbf{r}= x \hat{x}$  and $|\textbf{r}|=|x|$ denotes the distance between two atoms in 1D.
This function is simple because it is bounded and periodic.

On the other hand, the Green's function for free space is more complicated and given in Eq. (\ref{eqn:Greens_3d}):
\begin{equation}
    \begin{split}
         g(\mathbf{r}) = \text{i}\frac{\gamma_0}{2} &\bigg[h_0^{(1)}(k_0r)
         +\frac{3(\hat{r}\cdot\hat{q})(\hat{r}\cdot\hat{q}^*) - 1}{2} h_2^{(1)}(k_0r)
         \bigg]
         \label{eqn:Greens_3d}
    \end{split}
\end{equation} where $\hat{q}$ is the dipole orientation, $r = |\mathbf{r} |$ is the norm of the vector $\mathbf{r}$, $\hat{r} = \mathbf{r}/r$ is the unit vector along $\mathbf{r}$, and $h_l^{(1)}(x)$ are the outgoing spherical Hankel functions of angular momentum $l$; $h_0^{(1)}(x)=e^{ix}/[ix]$ and $h_2^{(1)}(x) = (-3i/x^3 - 3/x^2 + i/x)e^{ix}$. Unlike the 1D case, this Green's function is non-periodic. Its magnitude decays like $1/r$ at large $r$, and its real part diverges like $1/r^3$ at small $r$. In Eq. (\ref{eq:interaction_hamiltonian}), this leads to infinities in the self-interaction terms $i=j$, and we remove the real part of the Green's function at $i=j$ to avoid the infinities.


The distinction between $\textbf{R}_{i}$ and $\textbf{r}_{i}$ (see Fig. \ref{fig:system}) is important for the description of the motional aspect of the system but also has conceptual consequences for the following reason. The atoms interact through the electromagnetic field. As described in Eq. (\ref{eq:interaction_hamiltonian}) and in Eqs. (\ref{eq:1d_Greens_function} and \ref{eqn:Greens_3d}), the interaction strength depends on the phase difference of light due to the difference of absolute positions: $k_0|\tilde{\textbf{r}}_{ij}|$, where $k_0$ is the wave number of light. This has contribution from the relative positions of the traps, $k_0\textbf{R}_{ij}$, which is a fixed value. The other contribution comes from $k_0\textbf{r}_{ij}$, which is a quantum operator that describes the spatial degree of freedom of an atom inside its trap.

In the theoretical limit of infinitely trapped or infinitely massive atoms, this operator can be taken to be fixed, $k_0\textbf{r}_{ij} = 0$. This approximation is often used in theoretical studies of collective decay. For finite trapping strength, the position operator 
\begin{equation}
    k_0\textbf{r}_i = \eta\left(a_i + a_i^{\dagger}\right),
    \label{eq:vibrational_position}
\end{equation} where $\eta$ is the Lamb-Dicke (LD) parameter given by $\eta = k_0\sqrt{\frac{\hbar}{2 m \omega_t}}$. This LD parameter serves two purposes. First, it quantifies the coupling between the internal and vibrational states. The coupling $g(\tilde{\textbf{r}}_{ij})$ in Eq. (\ref{eq:interaction_hamiltonian}) can be Taylor expanded in the LD parameter to give terms where the vibrational operators, $a_i$ and $a_i^{\dagger}$, are multiplied by the internal operators $\sigma_{i}^{+}$ and $\sigma_{i}^{-}$. Second, the LD parameter quantifies the spatial spread of vibrational state of an individual atom. For example, the standard deviation of position  for the ground vibrational state divided by $\lambda_0$ is equal to ${\eta}/{2\pi}$. Typically, the LD regime is characterized by a spread that is much smaller than the wavelength, corresponding to $\eta \ll 2\pi$.

\subsection{Schrödinger equation}
\label{sec:schrodinger_equation}
The master equation in Eq. (\ref{eq:master_equation}) is computationally expensive to solve, especially for subradiant states which require substantial time to decay to the ground electronic state. In the case of starting in a pure state of the single excitation subspace with no driving laser, the master equation can be solved more efficiently in two parts: 1) a Schrödinger equation for the electronic excited state that does not depend on the electronic ground state dynamics. 2) a recycling superoperator that accumulates the decayed population in the electronic ground state.

In this paper, we always start the system from a pure state $\ket{\psi(t=0)}= \ket{DS,0}$, where $\ket{DS}$ is an electronic eigenstate and $\ket{0}$ is the vibrational ground state of all atoms. In this case, the total density matrix can be expanded as 
\begin{equation}
    \rho(t) = \rho_g(t) + \rho_e(t) = \rho_g(t) + \ket{\psi(t)} \bra{\psi(t)},
\end{equation} where $\rho_e(t) = \ket{\psi(t)} \bra{\psi(t)}$ is the pure excited state density matrix, and $\rho_g(t)$ is the decayed density matrix. The Schrödinger equation for the excited state is given by 
\begin{equation}
    \frac{d\ket{\psi(t)}}{dt} = \frac{-\text{i}}{\hbar}H_{\text{eff}}\ket{\psi(t)}.
    \label{eq:schrodinger_equation}
\end{equation} The magnitude of this state is equal to the probability of being in the excited state, $p(t)$.
\begin{equation}
    p(t) = \text{Tr}(\rho_e(t)) = \braket{\psi(t)}{\psi(t)}.
    \label{eq:probability}
\end{equation} This probability decays over time due to the non-hermiticity of the effective Hamiltonian. The rate of change of probability is given by
\begin{equation}
    \frac{dp(t)}{dt} =  \frac{2}{\hbar} \text{Im}\left[\bra{\psi(t)}\hat{H}_{\text{int}}\ket{\psi(t)}\right],
    \label{eq:prop_decay}
\end{equation} with a corresponding decay rate defined as 
\begin{equation}
    \gamma_d(t) = -\frac{dp(t)}{dt} / p(t).
\end{equation}
The decayed population $1-p(t)$ goes into the electronic ground state $\ket{g}\bra{g}$ where it evolves under the trap term.

\subsection{Observables}
\label{sec:observables}

To reconstruct the full density matrix. $\rho(t)$, we also need to solve for $\rho_g(t)$. However, keeping track of $\rho_g(t)$ is cumbersome, as it is usually a mixed state that evolves under the trap Hamiltonian.
\begin{equation}
    \frac{d \rho_g(t)}{d t} = -\frac{\text{i}}{\hbar} \left[H_{\text{trap}}\rho_g(t)-\rho_g(t)H_{\text{trap}}\right] + \mathcal{R}[\rho_e(t)].
\end{equation} Nevertheless, having the full $\rho(t)$ is rarely needed. We show how various quantities of interest such as the accumulated energy, infidelity, and entanglement entropy can be calculated from the evolved electronic excited state $\ket{\psi(t)}$ alone.

The accumulated vibrational energy at time t, $E_\text{R}(t) \equiv \langle H_{\text{trap}}\rangle (t)$, can be calculated by adding the energy of the electronic excited state and the electronic ground state $E_e(t)+E_g(t)$. $E_e(t)$ is equal to $\bra{\psi(t)}H_{\text{trap}}\ket{\psi(t)}$, and $E_g(t)$ is accumulated over time from the recycling superoperator:
\begin{equation}
    E_\text{R}(t) = E_e(t) + \int_{0}^{t} \text{Tr}\{H_{\text{trap}}\mathcal{R}[\ket{\psi(t)} \bra{\psi(t)}]\} dt.
    \label{eq:energy_recoil_time}
\end{equation} 

For highly subradiant configurations, storing a photon for a long time is of interest \cite{asenjo2017atom}. In particular, we wish to store the photon in a state of type $\ket{DS,0}$, where the electronic part is in a highly subradiant mode, $\ket{DS}$, and all atoms are in the vibrational ground state, $\ket{0}$. In the early dynamics of such a state ($t\ll1/\gamma_d$), the decayed population, $\text{Tr}(\rho_g(t))$, is small and the system is mostly in the excited state. 

However, due to the coupling between the electronic and vibrational states, the pure electronic eigenmode, $\ket{DS,0}$, generally turns into a superposition that includes other excited states that have different electronic or vibrational excitations. This leads to decoherence and infidelity in the stored excitation. The infidelity due to the mixing of the internally excited states is a measure of the closeness between the excited state at time $t$ and at time $0$ and is given by \cite{nielsen2010quantum}
\begin{equation}
    I(t) = 1 - \sqrt{\bra{DS,0}\rho^\text{proj}(t)\ket{DS,0}},
    \label{eq:infidelity}
\end{equation} where $\rho^\text{proj}(t) = \rho_e(t)/\text{Tr}(\rho_e(t))$. The reason we project the density matrix is to isolate the infidelity due to the mixing of the excited states from the infidelity due to the decay of the excited state. When the decayed population is small the infidelity due decay is $I_{\text{decay}}(t)\approx \text{Tr}(\rho_g(t))/2$. It is of interest to consider which infidelity dominates in early dynamics. In the context of conditional measurements, if there is no photon detected before time $t$, the state of the system is $\rho^\text{proj}(t)$ and only the infidelity due to the mixing of the excited states is relevant.

In general, the electronic and vibrational parts of the state $\rho_e(t)$ become entangled before the state decays. The entanglement entropy is calculated from the reduced density matrix of the internal states of the atoms $\rho_A(t) = \text{Tr}_{V}\left[\rho^\text{proj}(t)\right]$, where $\text{Tr}_{V}$ is the trace over the vibrational states of the atoms. The entanglement entropy is given by \cite{nielsen2010quantum}
\begin{equation}
    S(t) = -\text{Tr}\left[\rho_A(t)\text{ln}\left(\rho_A(t)\right)\right].
        \label{eq:entanglement_entropy}
\end{equation} One goal of this paper is to identify the conditions under which the entanglement entropy and the infidelity are minimized.

\section{Results}
\label{sec:results}

In this section, we show the effects of finite traps on the decay, recoil, and decoherence of dark states in quantum emitter arrays. We begin by studying two atoms in a 1D waveguide, where the atoms can be put in a highly subradiant configuration at $d = \lambda_0$ separation. We demonstrate that the decay rate can be significantly affected by the induced forces between the atoms. We calculate the energy imparted to the atoms during the decay and show that this energy could be much larger than the trap energy spacing, $\hbar \omega_t$, when the forces are significant.

We then extend our study to three atoms in a waveguide, where the initially separable system of electronic and vibrational states becomes entangled through the collective decay. We give numerical results and analytic approximations for the amount of infidelity and entropy generated with time. Finally, we extend this discussion to many-atom ring arrays in free space proposed in Ref. \cite{asenjo2017exponential}. We study the lower bound on the decay rate and explore qualitative relations for the dependence of the storage fidelity on the spread, trap frequency, distance between atoms, polarization of the electronic transition, and number of atoms.
\subsection{Two Atoms in a 1D Waveguide}
\label{sec:two_atoms_in_1d_waveguide}

We begin with the simplest example of highly subradiant states: two atoms in a 1D waveguide. In the single excitation subspace, the symmetrized combination can lead to subradiant or superradiant states when they are separated by a wavelength. The symmetrized electronic states are:
\begin{equation}
    \ket{\pm} = \frac{\ket{ge} \pm \ket{eg}}{\sqrt 2}.
    \label{eq:2_atoms_eigenstates}
\end{equation}
The effective Hamiltonian for this system is
\begin{align}
    H_{\text{eff}} &= \hbar \omega_t \sum_{i} a_i^{\dagger}a_i - \sum_{i} \frac{\text{i}\hbar \gamma_0}{2} \sigma_{i}^{+} \sigma_{i}^{-} \nonumber \\
    &\quad - \hbar g(\textbf{R}_{12}+ \textbf{r}_{12}) \left(\sigma_{1}^{+} \sigma_{2}^{-} + \sigma_{1}^{-} \sigma_{2}^{+}\right) \nonumber \\
    \label{eq:2_atoms_hamiltonian}
\end{align}

In the single excitation symmetrized states, the effective Hamiltonian acting on $\ket{\pm}$ becomes \cite{hybrid2025quantizedmotion}:
\begin{equation}
    H^{\pm}_{\text{eff}} = \hbar \omega_t \sum_{i} a_i^{\dagger}a_i - \frac{\text{i}\hbar \gamma_0}{2} \{1 \pm \exp\left(\text{i}\phi + \text{i}k_0\textbf{r}_{12}\right)\},
    \label{eq:2_atoms_hamiltonian_pm}
\end{equation} where $\phi = k_0 d$ is the phase difference due to the separation of the traps.

A highly subradiant configuration is achieved when the atoms are separated by a wavelength, $\lambda_0$, in the state $\ket{-}\ket{00}$. As shown in Ref. \cite{PhysRevLett.103.123903}, the initial decay rate of such configuration is given by
\begin{equation}
    \gamma_d(t=0) = 1- \exp(-\eta^2) \approx \gamma_0 \eta^2,
\end{equation}
where $\eta$ is the spread of the wavefunction, which is assumed to be small ($\eta \ll 1$) to give the quadratic scaling. In addition to this $t=0$ bound on the decay rate coming from non-zero spread, the later time $t>0$ decay dynamics of such configuration can be significantly affected as studied further in this section.

For identically trapped atoms in the $\ket{-}$ configuration, the effective Hamiltonian, $H^{-}_{\text{eff}}$, in Eq. (\ref{eq:2_atoms_hamiltonian_pm}) can be decomposed into two components: (1) a center-of-mass Hamiltonian, $H_{\text{cm}} = \hbar \omega_t a_{cm}^{\dagger}a_{cm}$, representing a pure harmonic oscillator with no perturbation, and (2) a relative motion Hamiltonian, $H_{\text{rel}}$, which is a harmonic oscillator perturbed by dipole-dipole interactions. At $d=\lambda_0$ separation, the relative Hamiltonian reads
\begin{align}
    H_{\text{rel}} = \hbar \omega_t a^{\dagger}a - \frac{\text{i}\hbar \gamma_0}{2} \{1 - \exp\left(\text{i}k_0\textbf{r}_{12}\right)\},
    \label{eq:relative_motion_hamiltonian}
\end{align}
where the non-indexed lowering operator, $a$, defined as $ a = (a_1-a_2)/\sqrt{2}$, acts on the relative displacement states. Here, $k_0 \textbf{r}_{12}$ can be expressed as $\sqrt{2} \eta (a + a^{\dagger})$.

The relative Hamiltonian in Eq. (\ref{eq:relative_motion_hamiltonian}) can be further expanded, assuming small $\eta$ and neglecting terms of order $\eta^3$ and higher, as
\begin{equation}
    \begin{split}
    H_{\text{rel}} &\approx \hbar \omega_t a^{\dagger}a + \frac{\hbar \gamma_0}{2} k_0 \textbf{r}_{12} - \frac{\text{i}\hbar \gamma_0}{4} (k_0 \textbf{r}_{12})^2 \\
    &= \hbar \omega_t a^{\dagger}a + \frac{\hbar \gamma_0 \eta}{\sqrt{2}} (a + a^{\dagger}) - \frac{\text{i}\hbar \gamma_0\eta^2}{2} (a + a^{\dagger})^2,
    \end{split}
    \label{eq:relative_motion_hamiltonian_expansion}
\end{equation}
The terms in this expansion represent the trap of strength $\omega_t$ (first term), a linear potential of strength $\gamma_0 \eta$ (second term), which pushes the wave packet away from the trap center, and non-uniform decay of strength $\gamma_0 \eta^2$ (third term), which deforms the wave packet because the wave function's tails decay faster than its center.
Throughout this work, we demonstrate that the relative strength of these terms determines the behavior of decay, recoil, and decoherence in the dark states. It is useful to consider three trap regimes: a \emph{strong trap} regime where the trap dominates the induced forces, $\omega_t\gg\gamma_0\eta$, a \emph{weak trap} regime where the induced forces dominate, $\omega_t\ll\gamma_0\eta$, and an \emph{intermediate trap} regime where the trap is only slightly stronger than the induced forces, $\omega_t\gtrapprox\gamma_0\eta$.



The effects of the induced forces and the non-uniform decay become manifest in weak traps. To isolate the effects of the different terms in (Eq. \ref{eq:relative_motion_hamiltonian_expansion}), we simulate two cases:
1) The full Hamiltonian as in Eq. (\ref{eq:relative_motion_hamiltonian}) with the induced forces present.
2) The Hamiltonian with the traps center shifted to cancel the linear potential term: $ H_{\text{rel}} \rightarrow H_{\text{rel}} -  {\hbar \gamma_0 \eta} (a + a^{\dagger}) /{\sqrt{2}}$. In other words, we cancel the linear part of the induced potential by adding a linear potential to the Hamiltonian. This removal of the \emph{linear} potential gives results very similar to simulating Eq. (\ref{eq:relative_motion_hamiltonian}) after deleting the \emph{full} coherent potential.

\begin{figure}[h]
    \includegraphics[width=0.45\textwidth]{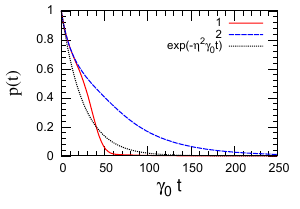}
    \caption{\justifying Decay of the excited population across cases discussed in the text. In both cases, the decay starts out in a similar way. An exponential decay at rate $\gamma_0 \eta^2$ is also plotted for reference.}
    \label{fig:decay}
\end{figure}

\begin{figure}[h]
    \centering
    \includegraphics[width=0.45\textwidth]{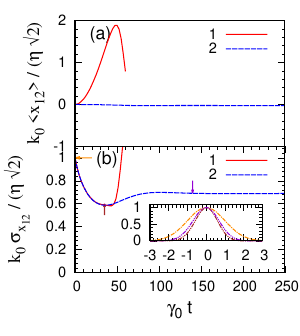}
    \caption{\justifying (a) the time-dependent expectation of the scaled relative position operator $k_0 x_{12} /(\sqrt{2}\eta)$ and (b) its standard deviation $k_0 \sigma_{x_{12}} /(\sqrt{2}\eta)$. The inset of (b) shows the time varying wavepacket probability for case 2 as a function of the of the scaled relative position, $k_0 x_{12} /(\sqrt{2}\eta)$. The curves (orange dash-dot, dark red dash-dot-dot, and purple dash-dot-dot-dot) correspond to three time instants ($\gamma_0t = 0,\, 35,\, 140$) respectively as indicated by the arrows. The red solid cureves (case 1) terminate at $t \approx 62/\gamma_0$, where the excited populations falls under 1\%.}
    \label{fig:motion}
\end{figure}

\begin{figure}[h]
    \centering
    \includegraphics[width=0.45\textwidth]{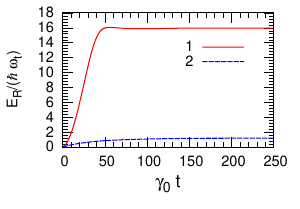}
    \caption{\justifying Energy gained by the atoms. In case 2 with shifted traps, the final energy is approximately $1.2 \hbar \omega_t$. In case 1 with forces present, the final energy substantially exceeds  $\hbar \omega_t$ and is about $16 \hbar \omega_t$.}
    \label{fig:recoil}
\end{figure}

The results of the simulations are in Figs. \ref{fig:decay}, \ref{fig:motion}, and \ref{fig:recoil}. In these simulations, we choose paramters that correspond to a Cesium atom transition $6^2 S_{1/2} \leftrightarrow 6^2 P_{3/2}$ with an experimentally relevant trap strength. The mass of the atom is $m = 2.21 \times 10^{-25}$ kg, and the individual atom decay rate is $\gamma_0 = 2\pi \times 5.2$ MHz. The internal transition has $\lambda_0 = 852$ nm, and the trap frequency is $\omega_t = 0.01 \gamma_0 = 2\pi \times 52$ KHz resulting in $\eta \approx 0.2$.  The system is initialized in the dark state configuration $\ket{-}\ket{00}$ with $\lambda_0$ separation.

In Fig. \ref{fig:decay}, we plot the time dependent excited state population for the two cases. Initially, both cases exhibit an exponential decay with a rate of approximately $\gamma_0 \eta^2$, but then deviate due to the change of the vibrational wavefunction. For the case where the induced force is balanced by shifted traps (case 2), the decay rate decreases. However, in the presence of forces (case 1), the decay rate greatly increases for $t>35/\gamma_0$.

The time dependence of the decay rate can be explained by the evolution of the spatial wavefunction. Figure \ref{fig:motion} plots the mean position and width of the wavefunction over time. With shifted traps (case 2), the wavepacket width decreases slightly due to faster decay of the wings away from the trap center (see the inset), while the position of the center of the wave packet remains nearly unchanged. This leads to a decrease in the standard deviation of position, $\sigma_{x_{12}}$, and thereby a decrease in the decay rate of Fig. \ref{fig:decay}. With forces present (case 1), the relative wavefunction shifts rightward, which is due to the repulsive interaction between the atoms’ individual wavepackets. The shift of the position of the wavefunction changes the relative phase $\phi$ from the initial value $2\pi$, resulting in a faster decay. The slowing and speeding of the decay discussed here are due to \emph{the motion or distortion} of the atomic wavepacket, but similar behavior can arise due to \emph{static disorder} of point-like atoms as in Ref. \cite{PhysRevA.109.013720}.


The motion of the wavepacket in Fig. \ref{fig:motion} not only affects the decay rate but also results in the atoms gaining kinetic and potential energy before they decay. This contributes to the excited state energy term, $E_e$, discussed in Sec. \ref{sec:observables}. This term builds up as the atoms are accelerated away from the trap center. This excited state energy is then transferred to the ground state energy term, $E_g$, as the atoms decay. When the atoms decay, they acquire additional recoil energy due to the emission of the photon. These two effects add up to the total energy of the atoms, $E_{\text{R}}$. As a consequence of these two contributions, the energy gained by the atoms exhibits qualitatively different behavior based on the trap strength. 

In the strong trap regime, $\omega_t\gg\gamma_0\eta$, the vibrational state experiences no motion, gaining minimal energy, $E_e(t)\approx 0$ for all $t$. In this case, the decay is exponential and the final energy is mostly from the recoil due to the emission of the photon. In this limit of a strong trap, the final vibrational energy of the atoms, $E_\text{R}$, is inversely proportional to the decay rate, as noted in Ref. \cite{suresh2021photon}, leading to
\begin{equation}
    E_{\text{R}}(\infty) \approx \frac{\gamma_0}{\gamma_d} E_{\text{r}},
    \label{eq:recoil_energy}
\end{equation}
where $E_{\text{r}} = \frac{\hbar^2 k_0^2}{2m}$ is the \emph{one-atom} recoil energy coming from momentum conservation during a photon emission from the state $\ket{e}$. The recoil energy in Eq. (\ref{eq:recoil_energy}) after photon emission is derived from the recycling term, assuming a stationary initial excited state. For the dark state of two atoms in a 1D waveguide, the decay rate is approximately $\gamma_0 \eta^2$, giving a recoil energy of 

\begin{equation}
    E_{\text{R}}(\infty) \approx \frac{E_{\text{r}}}{\eta^2} = \hbar \omega_t.
\end{equation} On average, a vibrational quantum is generated in this subradiant case in the strong trap regime, even in the LD regime. It is important to note that the effect of this recoil energy scales with the trap frequency and \emph{cannot} be suppressed by using a stronger trap. In contrast, the recoil energy from a single atom or subradiant states with decay coming mostly from geometric factors (such as two atoms with separation different from multiples of $\lambda_0/2$, or the 2D array configurations in in Ref. \cite{suresh2021photon}) is independent of the trap frequency (does not scale with $\omega_t$) and can be made insignificant by using a strong trap.

The motional energy gained by the atoms is shown for the two cases in Fig. \ref{fig:recoil}. When the traps are shifted to compensate for the induced force (case 2), the final energy is close to $\hbar \omega_t$, with difference ($E_R(\infty)-\hbar \omega_t$) coming from the narrowing of the wavepacket due to the non-uniform decay. However, with forces present (case 1), the final energy significantly exceeds $\hbar \omega_t$, mostly coming from the wavepacket displacement.

The energy gained by the atoms may be deposited either in the center-of-mass motion, $H_{\text{cm}}$, of the two atoms or in their relative motion, $H_{\text{rel}}$. For symmetric traps, a small fraction of the recoil energy, $E_{\text{r}}/2 = \eta^2 \hbar \omega_t/2$, is deposited in the center-of-mass mode.

The energy gained by the atoms and their decay behavior are manifest in the spectrum of the emitted photon \cite{loudon2000quantum}. The spectrum can be reconstructed following the `input-output' formalism as in Refs. \cite{gardiner1985input, caneva2015quantum, sheremet2023waveguide}. We calculated the spectrum of the emitted photon for the two atoms in a 1D waveguide in cases 1 and 2. In both cases the spectrum is red shifted on average by the final energy deposited in the atoms, $E_\text{R}(\infty)/\hbar$, and its width corresponds to the average decay rate of the subradiant state. The red shift accompanying the recoil might cause the output light to be off resonance with the atomic transition. This could affect applications where the output light is later used for manipulation of atomic states.
\subsection{Three Atoms in a 1D Waveguide}
\label{sec:3_atoms_in_1d_waveguide}

For a system of two atoms, the dark state \( \ket{-} \) is an eigenstate of the effective Hamiltonian. Consequently, this state remains decoupled from the bright state \( \ket{+} \) during the decay process, irrespective of atomic motion. This leads to the entanglement entropy \( S(t) \), as defined in Eq. (\ref{eq:entanglement_entropy}), remaining identically zero. However, at least two states are needed for a qubit. Therefore, we study a system of three atoms which can support two dark states.

In the single-excitation subspace, three atoms in a 1D waveguide, separated by $\lambda_0$, form two dark states and one bright state \cite{asenjo2017atom,paulisch2016universal,rubies2024deterministic}. The dark states can be defined as:
\begin{equation}
    \ket{\text{DS}_1} = \frac{1}{\sqrt{2}}\left(\ket{egg} - \ket{gge} \right),
    \label{eq:dark_state_1}
\end{equation}
\begin{equation}
    \ket{\text{DS}_2} = \frac{1}{\sqrt{6}}\left(\ket{egg} -2 \ket{geg} + \ket{gge}\right),
    \label{eq:dark_state_2}
\end{equation}
and the bright state is:
\begin{equation}
    \ket{\text{BS}} = \frac{1}{\sqrt{3}}\left(\ket{egg} + \ket{geg} + \ket{gge}\right).
    \label{eq:bright_state}
\end{equation}

If the atoms are fixed at integer wavelength separation, the dark states remain non-decaying while the bright state decays at a rate of $3 \gamma_0$ \cite{asenjo2017atom}. When the atoms have a non-zero spread in a vibrational ground state, the dark states initially decay at rate of $\simeq \eta^2 \gamma_0$ \cite{PhysRevLett.103.123903}, and the bright state decays at a rate of $\simeq (3-2\eta^2) \gamma_0$ in the Lamb-Dicke regime. Using the two dark states as qubit states allows high fidelity for a duration less than the lifetime ($ < 1/(\eta^2 \gamma_0)$). The lifetime is long because these subradiant states are almost exact eigenstates of the system and decay at the same, slow rate \cite{paulisch2016universal}.

\begin{figure}[h]
    \centering
    \resizebox{86mm}{!}{\includegraphics{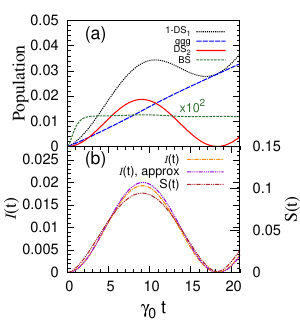}}
    \caption{\justifying Early dynamics of three atoms in a 1D waveguide. (a) Populations of all four internal states: the decayed population ($\ket{ggg}$), the population in $\ket{DS_2}$, the population in $\ket{BS}$ (multiplied by $10^2$ for clarity), as well as the initial state population $\ket{DS_1}$ (plotted as $1-DS_1$ to fit in the same range). (b) the infidelity $I(t)$ and the entropy $S(t)$ build up and oscillate as a proportional amount of $\ket{DS_1}$ is transferred back and forth to $\ket{DS_2}$. An analytical approximation, $2(\eta^2\gamma_0^2/\omega_0^2) \sin^2(\omega_t t/2)$, for $I(t)$ is derived in App. \ref{sec:3_atoms_state_interactions}. Chosen parameters correspond to $\omega_t = 0.34 \gamma_0$ and $\eta = 0.034$.}
    \label{fig:3_atom_dynamics}
\end{figure}

While the previous discussion focused on decay dynamics, practical traps introduce coupling between the two dark states due to atomic motion. For example, starting from \( \ket{DS_1,0} \), we observe population transfer to the electronic states \( \ket{DS_2} \) and \( \ket{BS} \), as shown in Fig. \ref{fig:3_atom_dynamics}. We use the same cesium parameters in Sec. \ref{sec:two_atoms_in_1d_waveguide}, except with a relatively stronger trap frequency, $\omega_t = 0.34 \gamma_0$, resulting in $\eta \approx 0.034$ . This choice ensures intermediate trapping conditions where the effects of the forces are small but still important to consider. When we simulated the weak trap in Sec. \ref{sec:two_atoms_in_1d_waveguide}, the strong induced forces quickly entangle the system, reaching maximum entropy. This is an unvaforable situation in information processing.

Unlike the complete decay dynamics in Fig. \ref{fig:decay} for two atoms, Fig. \ref{fig:3_atom_dynamics} only shows the early dynamics where the decayed population, $1-p(t)$, is under $3\%$. For reference, if the atoms are confined to the ground state of the trap, the lifetime $1/\gamma_d$ is approximately $1/(\eta^2\gamma_0)=865/\gamma_0$, while we only simulate for $t_f = 21/\gamma_0$. In this time window, the state $\ket{DS_2}$ (summed over all vibrational states) populates up to approximately $2\%$ and oscillates with a period close to the trap period $2\pi/\omega_t \approx 18.5/\gamma_0$. The bright state $\ket{BS}$ also populates to a small fraction $\approx 1 \times 10^{-4}$. The simple sinusoidal behavior results because the dipole-dipole interaction couples \( \ket{DS_1,0} \) with $\ket{DS_2}$ in the $n=1$ subspace, and these states have an energy splitting $\simeq \hbar \omega_t$, see App. \ref{sec:3_atoms_state_interactions}.

This mixing between the states is only possible owing to the coupling between the internal and the vibrational states. In the process of the population transfer from the initial state, there is entanglement entropy ($S(t)$ as defined in Eq. (\ref{eq:entanglement_entropy})) generated. The entropy is approximately proportional to the population in $\ket{DS_2}$. For example there is a peak in the entropy at $t \approx 9.2/\gamma_0$ which corresponds to the time when the population in $\ket{DS_2}$ is at its maximum. This peak is at $S(t) \approx 0.1$ which is small but could affect applications. Like the entropy, populating $\ket{DS_2}$ causes infidelity, $I(t)$, that is also proportional to the population in $\ket{DS_2}$ and peaks at a value approximately $0.02$. Note that the $I(t)$ is correlated with $S(t)$, as they both depend on the population in $\ket{DS_2}$. This peak mixing infidelity is of the same order of magnitude as the decay infidelity, which signifies the importance of including the effects of motion for studying the quality of subradiant states in early time.

The oscillation amplitude for the entropy or the infidelity curves can serve as a figure of merit to compare the effects of the trap strength and the induced forces across different system parameters, $\eta$ and $\omega_t$. In the intermediate or strong trap regime, we found that the amplitude of the oscillations depends on the ratio of the force term to the trap term, similar to the condition for exponential decay for two atoms. Specifically, the infidelity amplitude scales as $\sim (\gamma_0 \eta / \omega_t)^2$. This scaling, derived in Appendix \ref{sec:3_atoms_state_interactions}, arises because the force term, $\gamma_0 \eta$, functions as a coupling in an effective two-level system, while the trap term, $\omega_t$, serves as a detuning. 


\subsection{Circular Array of Atoms in Free Space}
\label{sec:circular_array_of_atoms_in_free_space}

In this section, we present results for atoms interacting in free space, without a waveguide. To achieve highly subradiant configurations in free space, we arrange multiple atoms in a regular array with separation $d < 0.5 \lambda_0$ \cite{asenjo2017exponential}. For a large number, $N$, of point-like atoms arranged in a circular array, the decay rate of the most subradiant eigenstate can be made very small by increasing $N$. In the infinite limit $N\rightarrow\infty$, the decay rates of many eigenstates approach zero. These eigenstates are electronic (spin) waves with different wavenumbers $k$. When $k$ exceeds the free-space wavenumber $k_0$, the radiation must be evanescent perpendicular to the array, resulting in guided modes around the ring that are completely dark. For finite even $N$, the most subradiant eigenmode is the one with the largest $k$ which is equal to $\pi/d$. This state which we call $\ket{DS_{min}}$ is an equal superposition of all single excitation states with alternating signs. The decay rate of this state, $\gamma_{min} = 2\,\text{Im}\{\bra{DS_{min}}H_{\text{eff}}\ket{DS_{min}}\}/\hbar$, decreases exponentially as $\exp(-N/N_0)\gamma_0$, where $N_0$ is a constant that depends on the interatom distance, $d$. The resulting highly subradiant states of the ring are promising for the storage of photons and quantum information \cite{PhysRevResearch.4.013110}.

If the atoms start in a Gaussian profile of non-zero spread in a finite trapping potential, as shown in Fig.~\ref{fig:system}, the initial decay rate of the most subradiant state, $\gamma_{min}(t=0)$, is equal to $2\,\text{Im}\{\bra{DS_{min},0}H_{\text{eff}}\ket{DS_{min},0}\}/\hbar$. At later time, the dynamics might shift or deform the vibrational wavefunction resulting in a modified decay rate. $\gamma_{min}$ has two primary contributions: (1) the finite size of the array with afromentioned exponential improvement with $N$, and (2) the finite spread of the atoms' wavefunction, which contributes a factor of $\sim \eta^2 \gamma_0$ \cite{PhysRevLett.103.123903,guimond2019subradiant}. At sufficiently large $N$ and in the LD regime, $\gamma_{min}$ reaches an asymptotic value dominated by the atomic spread, $\gamma_{spread}= C \eta^2 \gamma_0$, where $C$ is a proportionality constant. The value of this proportionality constant has an interesting behavior. First, it depends on the istoropy of the atomic spread and the polarization of the transition but is \emph{independent of the interatom distance $d$} for all $d < 0.5 \lambda_0$. Second, for atoms with isotropic spread, $C = 1$ independent of polarization (see appendix G of \cite{PhysRevA.111.053712}). For an atom ring laid in the x-y plane with uniform x-y spread but no spread in the z direction, the value of $C$ depends on the polarization direction. For linear polarization perpendicular to the plane of the atoms ($\hat{q}=\hat{z}$), $C \approx 0.8$, while for polarization in the x-y plane ($\hat{q} \perp \hat{z}$), $C \approx 0.6$. On the other hand, for spread in z twice that in the x-y plane, $C \approx 1.6$ for $\hat{q}=\hat{z}$. For $\hat{q} \perp \hat{z}$, $C \approx 2.2$. More discussion on the asymptotic decay rate and the proportionality constant can be found in Appendix \ref{sec:equations_for_circular_array}.

\begin{figure}[h]
    \resizebox{86mm}{!}{\includegraphics{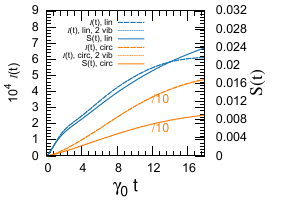}}
    \caption{\justifying Similar to Fig.~\ref{fig:3_atom_dynamics}(b), the evolution of the most subradiant internal eigenstate for $N$ atoms in a ring configuration, with motion initialized in the ground state of the trap, $\ket{DS_{min},n=0}$. Population transfers from the initial internal state to other internal states in the first vibrational mode $n=1$ (not shown), leading to increasing infidelity and entropy between the evolved state and the initial state, $I(t)$. Parameters: $\omega_t = 0.34 \gamma_0$, $\eta = 0.034$, $d = 0.3 \lambda_0$, $N= 30$, linear, $\hat{q}=\hat{z}$, or circular polarization, $\hat{q}=(\hat{x}+\text{i}\hat{y})/\sqrt{2}$. Note that the curves for circular polarization are divided by 10 for clarity. All curves, except where indicated, use one vibration restricted Hilbert space.}
    \label{fig:entanglement_ring}
\end{figure}

Similar to the case of three atoms, forces between atoms in the ring induce entanglement between their internal and vibrational degrees of freedom. Figure~\ref{fig:entanglement_ring} shows the mixing dynamics starting from $\ket{DS_{min},0}$. The frequency used here is the same as in Sec. \ref{sec:3_atoms_in_1d_waveguide}, $\omega_t = 0.34 \gamma_0$, corresponding to intermidiate trapping, with $d = 0.3 \lambda_0$, $N = 30$, and linear polarization, $\hat{q}=\hat{z}$ (circular polarization, $\hat{q}=(\hat{x}+\text{i}\hat{y})/\sqrt{2}$).  In the time shown, approximately $0.6\%$ ($1\%$) of the population has decayed to the ground state $\ket{gg \dots g}$. A small fraction, about $0.1\%$ ($1\%$), transfers from the initial internal state to other internal states in the first vibrational mode, $n=1$, (summed over the all the internal states). This process generates entanglement entropy $S(t)$ and produces infidelity $I(t)$ as shown in Fig. ~\ref{fig:entanglement_ring}. Because the infidelity and entropy are both mainly caused by mixing with $n=1$ manifold, they are correlated; i.e, $S(t)$ increases as $I(t)$ increases and vice versa.

Compared to the 3 atoms case in Fig.~\ref{fig:3_atom_dynamics} which has the same $\eta$ and $\omega_t$, the infidelity here is much smaller, of order $10^{-4}$ ($10^{-3}$), compared to $10^{-2}$ for the waveguide atoms. This can be explained by the magnitude of the induced forces between the atoms. For a waveguide at $\lambda_0$ separation, the force is $\hbar\gamma_0k_0$. For free space at $d = 0.3 \lambda_0$ and when the polarization is perpendicular to the plane of the atoms, the force is much smaller, approximately $0.03\hbar\gamma_0k_0$. This is a particularly useful arrangement for storing quantum information, as the decay rate is small and the induced forces are weak. The force stays small for $d/\lambda_0$ in the range from approximately $0.25 $ to $0.5$ but increases rapidly for smaller $d$, as later indicated in Fig.~\ref{fig:infidelity_vs_d_ring}. For free space at $d = 0.3 \lambda_0$ and circular polarization, the force is larger than the linear case and is approximately $0.64\hbar\gamma_0k_0$.  This results in the population leaking to the other excited states ($1\%$) being comparable to the decayed population ($1\%$), suggesting that the mixing effects on the overall fidelity due to vibrations are as significant as the spontanteous decay process.

Unlike the periodic mixing seen for three atoms in Fig.~\ref{fig:3_atom_dynamics}, the mixing here is not periodic and generally more complex at early times. The lack of periodicity can be explained by the change of energy splitting between the interacting states. For atoms in a waveguide, the splitting is $\omega_t$, which causes oscillations with frequency close to $\omega_t$. While being absent in a waveguide at $d=\lambda_0$, another contribution to the splitting comes from the Lamb shift due the collective coupling in free space at $d = 0.3 \lambda_0$.

To provide qualitative explanation for the observed dynamics, we move to the Bloch basis for both internal and vibrational states labeled by wavenumbers $k$ and $k_{vib}$, respectively. Because of the symmetry of the ring configuration, the \emph{sum} of these quasi-momenta is conserved to yield a non-zero interaction with the initial state $\ket{DS_{min},0}$. We label the interaction matrix element as $\mathcal{G}_{k}$ which is usually non-zero after $k_{vib}$ is chosen to satisfy the conservation. In the intermediate trap regime($\eta \gamma_0 \lesssim \omega_t \lesssim \gamma_0$), at most one phonon can be considered. The initial state, which lies in the  0-phonon ($n = 0$) band, interacts weakly with the 1-phonon bands with strength $\mathcal{G}_{k}$ that varies with $k$ and scales as $\eta \gamma_0$. The initial state has an energy shift and decay rate captured by a complex matrix element $\mathcal{E}_{min} = \bra{DS_{min},0}H_{\text{eff}}\ket{DS_{min},0} \equiv \hbar(\Delta_{min}-i \gamma_{min}/2)$. Similarly, the 1-phonon states have complex energy $\mathcal{E}_{k} + \hbar\omega_t$. The shape of such bands depends on the interatom distance as well as the polarization direction and can be found in \cite{asenjo2017exponential}. By inspecting the effective interaction strength $S(k) = |{\mathcal{G}_{k}/(\mathcal{E}_{k} + \hbar\omega_t - \mathcal{E}_{min})}|^2$, we found that a group of only 6 states interact strongly with the initial state. These states come into two pairs that are separated in $k$, but nevertheless have similar energy ($\mathcal{E}_{k}$). The analysis was done for the cases in Fig.~\ref{fig:entanglement_ring}, and for an increasing number of atoms, $N=60$ and $120$. The narrow bandwidth of the interacting ($n=1$) states motivates modeling the system as an effective 2-level system which captures the qualitative features of the infidelity dyanmics.

For example, the global maximum of the mixing infidelity in Fig. ~\ref{fig:entanglement_ring} is proportional to the largest $S(k)$. Additionally, the frequency of the oscillation and the revival time of the fidelity are determined by the energy splitting (real part of the denominator). In Fig. ~\ref{fig:entanglement_ring}, the Lamb shift in the linear polarization case is approximately $\Delta_{min} = -0.25 \gamma_0$ for the initial state and $-0.55 \gamma_0$ for the strongest interacting state. This uneven Lamb shift causes the energy splitting between the interacting states to diminish from $\omega_t = 0.34\gamma_0$ (the trap splitting) to $\approx 0.04\gamma_0$. This results in a slower oscillation and a revival time longer than $\sim 1/\omega_t$ and the time window of interest. While the 2-level model can capture the time scales for the infidelity revival or saturation, it tends to underestimate the peak of the infidelity. For example, in both cases in Fig.~\ref{fig:entanglement_ring}, the peak is underestimated by a factor of $\sim 2$. This is because other less strongly interacting states were ignored. Nevertheless, this qualitative picture provides a way to considerably minimize the mixing the infidelity. This is achievable by increasing the trap frequency to be larger than the width of the energy band, which is usually of order of a few $\gamma_0$.

Due to the increasing size of the Hilbert space with $N$, the simulation in Fig. ~\ref{fig:entanglement_ring} uses a restricted vibrational Hilbert space allowing at most one atom to vibrate in one direction at a time. This restriction was validated by comparison to another restriction with at most two vibrations. The errors in populations and infidelity in the time window shown in Fig. ~\ref{fig:entanglement_ring} between one and two vibrations were less than $1\%$. The errors were even smaller for cases with larger $\omega_t$ or smaller $\eta$. For times longer than shown in Fig. ~\ref{fig:entanglement_ring}, the infidelity for the circular polarization case increases to the 1\% level and the vibrational restriction is no longer convergent (relative error between one- and two-vibration restrictions is $>1\%$). This suggests that such restriction results in a good approximation only when the system is in the intermediate trap regime, or when the evolution time is early enough that the $n=1$ population has not increased significantly. Further details of the equations and approximations used are given in the App. \ref{sec:equations_for_circular_array}.

\subsubsection{Trends with Different Parameters}
\label{sec:trends_with_different_parameters_ring}

In this section, we find the optimal parameters for the ring configuration that minimize the vibrational effects. To achieve that, we analyze how infidelity depends on various parameters \((N, d, \eta, \omega_t, \hat{q} )\). First, we time-evolve the excited state \(\ket{DS_{min},0}\) till a significant fraction of it (which we take to be $2\%$) decays to the electronic ground state. This threshold is chosen to ensure that the electronic excitation is still preserved with high probability. We later discuss other choices for the threshold ($1\%$, $4\%$, and $8\%$).
In this time window, we calculate the mixing infidelity, $I(t)$, in Eq. (\ref{eq:infidelity}) and record its maximum value, which we denote as \(I_{\text{max}}\). This maximum infidelity is then compared across various parameters. Additionally, we compare these infidelity values coming from the mixing in the excited states to the infidelity coming from decay ($I_{\text{decay}} \approx \text{Tr}(\rho_g)/2=0.01$).

Consistent with previous findings in waveguide-coupled atoms, the infidelity exhibits a quadratic dependence on \(\eta\) for sufficiently small \(\eta\) and decreases as \( (\omega_t / \gamma_0)^{-2} \) for sufficiently large \(\omega_t\). However, as \( N \) varies, the infidelity remains within the same order of magnitude for sufficiently large \(N\), displaying only minor fluctuations.

\begin{figure}[t]
    \centering
    \includegraphics[width=0.45\textwidth]{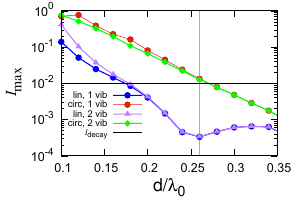}
    \caption{\justifying Maximum infidelity, $I_{\text{max}}$, versus interatom separation, $d$, for linear polarization, $\hat{q}=\hat{z}$, and circular polarization, $\hat{q}=(\hat{x}+\text{i}\hat{y})/\sqrt{2}$. Parameters: $\omega_t = 0.34 \gamma_0$, $\eta = 0.034$, $N= 30$. These infidelities from the mixing of the excited states are compared to the decay infidelity indicated by the horizontal line. To the right of the vertical orange line, the configuration is \emph{not} in the spread dominated decay. We use the one-vibration restriction and validate it with a two-vibration restriction as described in the text.}
    \label{fig:infidelity_vs_d_ring}
\end{figure}

As the interatomic distance \( d \) decreases, there are stronger induced forces, leading to increased infidelity, as illustrated in Fig.~\ref{fig:infidelity_vs_d_ring}. Notably, the magnitude of induced forces depends on the polarization direction. For instance, in the case of linear polarization along \( \hat{z} \), the induced forces are weaker than those observed for circular polarization along \( \hat{q} = (\hat{x} + \text{i} \hat{y})/\sqrt{2} \). Consequently, the infidelity is significantly higher for the circular polarization case. At \( d = 0.3 \lambda_0 \), for example, the infidelity is on the order of \( 10^{-4} \) for perpendicular polarization, whereas it increases to \( 10^{-2} \) for circular polarization. Compared to the decay infidelity, the mixing infidelity is much smaller for perpendicular polarization and only becomes important at small distances $d \lesssim 0.13$. On the other hand, for circular polarization the mixing infidelity is much larger than that of the decay and only becomes insignificant at bigger values of $d$.

At larger values of inter-atom separation, the mixing infidelity, $I_{\text{max}}$, generally decreases with increasing $d$, as the induced forces become weaker. For  ciruclar polarization, there is a sharp drop for the infidelity with larger $d$. The reason is that the used number of atoms, $N=30$, is not large enough to reach the asymptotic limit of the decay rate, $\gamma_{spread}$, for $d > 0.26 \lambda_0 $, indicated by the vertical line. For $d > 0.26 \lambda_0 $, the system decays faster with larger $d$. Consequently, the infidelity does not have enough time to grow to the same level as for smaller $d$. This is especially observed if the chosen threshold is increased or decreased from $2\%$. For linear polarization, the infidelity at small $d$ reaches its maximum and saturates even before the $1\%$ decay time. At $d > 0.24 \lambda_0$, the infidelity is still growing and only reaches saturation at a later time. For linear polarization, all convergent points (with respect to number of vibration allowed) saturate at the $2\%$ threshold except for the largest $d$ which saturates at $4\%$. For circular polarization, the infidelity saturates at the $4\%$ or $8\%$ threshold for convergent points. Nevertheless, the infidelity only grows \emph{within} the same order of magnitude shown in Fig.~\ref{fig:infidelity_vs_d_ring}, and thus the $2\%$ threshold suffices to capture the qualitative features.

The above analysis suggests optimal operating parameters for the ring configuration. First, perpendicular polarization is preferable to circular polarization because the vibrational effects are insignificant compared to the decay for a bigger range of $d$. If circular polarization is to be used, larger values of $d$ are needed to minimize the vibrational effects. Subsequently, more atoms ($>30$ in the above case) are needed to reach the asymptotic spread decay.

The analysis in this section is performed under the same restriction in Fig. \ref{fig:entanglement_ring}: at most one direction of one atom is in the $n=1$ state at a time, which we validate using a two-vibration restriction. At large values of $d$, the forces are small resulting in the $n=1$ population being relatively small. In this case, the one-vibration agrees well with the two-vibration restrictions as seen in Fig.~\ref{fig:infidelity_vs_d_ring}. At smaller $d$ where the induced forces are large, there is a mismatch between the one- and two-vibration restrictions. Consequently, more vibrations need to be included in the simulations to get accurate results. Nevertheless, high infidelity situations are not expected to be relevant for practical applications.


\section{Conclusion}
\label{sec:conclusion}
In this work we expanded on the calculations of the decay mechanism and the recoil energy introduced in Refs. \cite{suresh2021photon,suresh2022atom} in the limit of highly subradiant states. In the strong trap regime, the recoil energy is equal to one harmonic energy level separation, $\hbar \omega_t$, which is concentrated mostly in the relative vibrational mode of the two atoms. This recoil might cause severe heating of the atoms that would require additional cooling steps each time a photon is emitted from a dark state of an atomic array \cite{rubies2024collectively}.

We also showed that weakly trapped atoms can suffer from the induced forces during the decay. The forces can have detrimental effects by either deforming or moving the motional wavefunction from the dark state configuration leading to accelerated decay at later time $t>0$. In this case, the atoms receive significant amounts of vibrational energy from the induced forces during the decay. In a 1D waveguide, these could be orders of magnitude bigger than the harmonic level separation. To quantify when the induced forces become important, we showed that for a 1D waveguide a large ratio of the trap to the force, $\omega_t/\eta\gamma_0$, is required to maintain the exponential decay and alleviate the effects of the forces. This can be achieved by either having stronger traps or more massive emitters.

Finally, we analyzed the effect of motion on the quality of the subradiant states in free space. We explored the limiting spread decay for an atoms in a ring configuration, which is proportional to $\eta^2 \gamma_0$ for initial Gaussian wavefunctions at $t=0$ \cite{PhysRevLett.103.123903, guimond2019subradiant}, and analyzed its possible dependence on interatom separation, polarization of the transition, and isotropy of the spread. Furthermore, we explored the infidelity and entanglement entropy due to the induced forces. These quantities can be suppressed by a stronger trap or choice of certain polarizations and inter-atom distances that minimize the forces. These results have implications on the design of quantum memories that store photons in subradiant states \cite{PhysRevResearch.4.013110} and the design of atom array experiments.

Data plotted in the figures is available at \cite{data}.

\begin{acknowledgments}
    This work was supported by the National Science
    Foundation under Award No. 2109987-PHY and 2410890-PHY. We thank Deepak Suresh for helpful discussions.
\end{acknowledgments}
\appendix



\section{3 atoms state interactions}
\label{sec:3_atoms_state_interactions}

In this appendix, we derive a scaling relation for the decoherence of the atoms in the 1D waveguide assuming an intermediate trapping regime, where the trap is strong enough to limit the vibrational excitations to $n_{max}=1$. For atoms with equal masses and trap frequency and separation of $\lambda_0$, the system possesses a parity symmetry that simplifies the dynamics of the the Hamiltonian in Eq. (\ref{eq:interaction_hamiltonian}). This parity symmetry is realized by reflecting the atoms (both electronic and vibrational) around the center of the array. For $N$ atoms, this is realized by swapping the electronic operators $\sigma_i \leftrightarrow \sigma_{N-i}$ and the vibrational operators $\mathbf{r}_i \leftrightarrow -\mathbf{r}_{N-j}$. This symmetry leaves the Hamiltonian invariant, and as a result, the Hilbert space can be divided into two sectors that do not interact, the sectors with odd and even parity of the combined internal and vibrational degrees of freedom.

For 3 atoms in the $n_{max}=1$ approximation, there are 4 allowed vibrational states $\ket{n_1n_2n_3} \in \{\ket{000},\ket{100},\ket{010},\ket{001}\}$. The other higher vibrational states can be ignored because their coupling to the initial state is of order $\mathcal{O}(\eta^2)$ which is much smaller than the the coupling to the one-vibration states (of order $\mathcal{O}(\eta)$). The one-vibration states can be symmetrized just like for the electronic states.

The symmetrized states are given by
\begin{equation}
    \begin{split}
        \ket{0} &= \ket{000} \\
        \ket{V_1} &= \frac{1}{\sqrt{3}} (\ket{100} + \ket{010} + \ket{001}), \\
        \ket{V_2} &= \frac{1}{\sqrt{6}} ( \ket{100} - 2\ket{010} + \ket{001}), \\
        \ket{V_3} &= \frac{1}{\sqrt{2}} (\ket{100} - \ket{001}).
    \end{split}
    \label{eq:symmetrized_states}
\end{equation}With respect to the spatial parity transformation, $\mathbf{r}_i \leftrightarrow -\mathbf{r}_{N-j}$, the states $\ket{0}$ and $\ket{V_3}$ have even parity while the states $\ket{V_1}$ and $\ket{V_1}$ have odd parity. This can be achieved by applying this transformation on the product states. For example, the state $\ket{001}$ is sent to $-\ket{100}$. In this way, the full Hilbert space breaks into two subspaces. Because we start in the $\ket{DS_1}\ket{0}$ state, the relevant sector is spanned by $\{\ket{DS_1}\ket{0},\ket{DS_1}\ket{V_3},\ket{DS_2}\ket{V_1},\ket{DS_2}\ket{V_2},\ket{BS}\ket{V_1},\\ \ket{BS}\ket{V_2}\}$ which have an odd parity of the full system symmetry. The other states have even parity and do not couple to the initial state.

In our case, there is another simplifying symmetry. Since the forces are pairwise equal in magnitude and opposite in directions, the net force on all atoms is zero resulting in the center of mass vibrational mode ($\ket{V_1}$) separating from the relative modes ($\ket{V_2}$ and $\ket{V_3}$). In this way, the dynamics of the odd sector could be reduced to a Hamiltonian which in the $\{\ket{DS_1}\ket{0},\ket{DS_1}\ket{V_3},\ket{DS_2}\ket{V_2},\ket{BS}\ket{V_2}\}$ basis is (to $\eta$ order)

\begin{equation}
    \frac{H_{\text{odd}}}{\hbar} = 
    \begin{pmatrix}
        0 & \frac{\eta  \gamma_0}{\sqrt 2} & \frac{\eta  \gamma_0}{\sqrt 2} & \frac{-\eta  \gamma_0}{2} \\
        \frac{\eta  \gamma_0}{\sqrt 2} & \omega_t & 0 & 0 \\
        \frac{\eta  \gamma_0}{\sqrt 2} & 0 & \omega_t & 0 \\
        \frac{-\eta  \gamma_0}{2} & 0 & 0 & \omega_t -\frac{3\text{i}  \gamma_0}{2}
    \end{pmatrix}.
    \label{eq:H_odd}
\end{equation} Note that the terms responsible for the decay of the subradiant states are omitted because they are of order ($\eta^2$). Because we are interested in the early time dynamics, the decayed population is small and can be ignored.

A simpler model can be obtained from ignoring the bright state that only populates in a tiny fraction and decays at fast rate as in Fig. \ref{fig:3_atom_dynamics}. The other states $\ket{DS_1}\ket{V_3},\ket{DS_2}\ket{V_2}$ behave symmetrically with respect to the $\ket{DS_1}\ket{0}$ and can be further symmetrized to obtain the 2 level Hamiltonian in the $\{\ket{DS_1}\ket{0},\ket{S}\}$ basis where $\ket{S}$ is the symmetric superposition of $\ket{DS_1}\ket{V_3}$ and $\ket{DS_2}\ket{V_2}$.
\begin{equation}
    \frac{H_{\text{2-level}}}{\hbar} = 
    \begin{pmatrix}
        0 & \eta \gamma_0 \\
        \eta \gamma_0 & \omega_t 
    \end{pmatrix}.
    \label{eq:H_2_level}
\end{equation} This simple Hamiltonian captures the essential entanglement dynamics in early time. It resembles the Rabi oscillation model where $\omega_t$ is the detuning, and $\eta \gamma_0$ is the driving Rabi frequency. When the driving Rabi frequency is small the population of the exited state in early time can be found analytically to be
\begin{equation}
    P_S(t) \approx 4 \frac{\eta^2 \gamma_0^2}{\omega_t^2}\sin^2(\frac{\Omega t}{2}),
\end{equation} where $\Omega = \sqrt{\eta^2 \gamma_0^2 + \omega_t^2}$ is the beating frequency. $\Omega$ can be approximated as $\omega_t$ for small $\eta$, in the intermediate or strong trap regime. Since the state $\ket{S}$ has equal amounts of $DS_1$ and $DS_2$ that are in different vibration modes, the system becomes entangled. The amount of population in $\ket{DS_2}$ is given by $P_{DS_2}(t) = P_S(t)/2$.

The resulting fidelity is approximately equal to the classical population fidelity $\sum_i \sqrt{p_i q_i}$ for two distributions $p_i$ and $q_i$. In our case we have two states starting from $p_1=1,p_2=0$ and at time $t$ the populations become $q_1=1-q_2,q_2=P_{S}(t)$. The resulting fidelity is thus $\sqrt{1-P_{S}(t)}$ in early time. For small excited population, the infidelity becomes approximately $P_{S}(t)/2$ which is the case for the simulation in fig. \ref{fig:3_atom_dynamics}, where the infidelity is close to $2\%$ at its peak. The resulting entanglement entropy is also approximately proportional to $P_S(t)$, although with a different proportionality factor.

\section{Equations and approximations for a circular array of atoms}
\label{sec:equations_for_circular_array}

In this appendix, we analyze the atoms on the ring. The ring is assumed to be in the $x$-$y$ plane with atoms polarized linearly or circularly in $z$ direction. The locations of the centers of the traps are at $\textbf{R}_j=\langle X_j, Y_j \rangle = \langle r\cos(j\theta), r\sin(j\theta)\rangle$, where $\theta = 2\pi/N$ and the radius $r$ is equal to $d/(2\sin(\theta/2))$. 

Under intermediate trapping conditions, the forces are strong enough to populate the first vibrationally excited state, $n=1$, but not strong enough to populate higher vibrationally excited states. Therefore, we obtain a computationally feasible Hilbert space by restricting the vibrations to at most one atom vibrating in one direction at a time. In this case, a vibrationally excited $\ket{V} \in \{a_{l,i}^{\dagger}\ket{0}\}$, where $i$ is an index for the atoms and $l$ is either $x$, $y$, or $z$. Because the induced force is colinear with the atom-atom separation line, the out-of-plane $z$ componenet will be suppressed by $\eta/(k_0 d)$ factor compared to the in-plane $x-y$ component, and is simply ignored in the following analysis. The one-vibration restriction breaks down when the induced force is large, but helps to qualitatively estimate its effect. For the following analysis, the atoms have infinite trap and zero spread in the $z$-direction, but we then analyze the effect of inclusion of the $z$ spread. 

With either polarization described above, the Green's function in Eq. (\ref{eqn:Greens_3d}) simplifies to a function of the phase: $g(\textbf{r}) = f(\phi)$, where $\phi = k_0 r = k_0 \sqrt{x^2+y^2}$. The decay rate and effect of the induced force can be estimated by analyzing the relevant matrix elements of the interaction Hamiltonian in the LD regime. The initial decay rate (and the Lamb shift) can be computed from the matrix elements:

\begin{equation}
    \begin{split}
        &\bra{g,0}\sigma_i^{-}H_\text{int}\sigma_j^{+}\ket{g,0} \\ 
        &=- \hbar \left\{f(\phi_{ij}) + \eta^2 \left[f''(\phi_{ij}) + \frac{f'(\phi_{ij})}{\phi_{ij}}\right]\right\} , \\
    \end{split}
    \label{eq:matrix_elements_00}
\end{equation} for the ground vibrational states. Here, $\phi_{ij}=k_0 \sqrt{(X_i-X_j)^2+(Y_i-Y_j)^2}$. Note there is an exception to the above rule for $i=j$, in which case the matrix element is equal to $-\text{i}\hbar\gamma_0/2$. There is a contribution from the point-like atoms (the first term) and the spread of the atoms of order $\eta^2$ (the second term). The force on the atoms is related to the matrix elements connecting the ground and first vibrational states which are given by

\begin{equation}
    \begin{split}
        &\bra{g,0}\sigma_i^{-}a_{x,k}H_\text{int}\sigma_j^{+}\ket{g,0} \\
        &= - \hbar \eta \frac{k_0X_{ij}f'(\phi_{ij})}{\phi_{ij}} (\delta_{ik}-\delta_{jk}), \\
    \end{split}
    \label{eq:matrix_elements_01}
\end{equation} and similarly for the $y$ direction. The form of the function $f$ is relatively complicated, so we calculate the derivatives numerically using finite differences.

The system of the atoms on the ring exhibits a high degree of symmetry (Dihedral group of order $N$). As a result, it can be shown that the eigenstates of the system are spin waves when restricting to the $n=0$ subspace \cite{antman2024atom}. For example, for even $N$ the most subradiant state $\ket{DS_{min}}$ is a spin wave with wave number $k_{min} = \pi/d$. 
\begin{equation}
    \ket{DS_{min}} = \frac{1}{\sqrt{N}}\sum_{j=1}^{N} \exp(\text{i}k_{min}jd)\sigma_j^{+}\ket{g,0}.
    \label{eq:DS_min}
\end{equation}
The eigenvalue associated with this state results in the minimum decay rate of the system.
\begin{equation}
    \begin{split}
        \gamma_{min} &= \frac{2}{\hbar}\, \text{Im} \bra{DS_{min}}H_{\text{int}}\ket{DS_{min}} \\
            &= 2 \sum_{j=1}^{N} (-1)^j \text{Im}\left\{f(\phi_{1j}) + \eta^2 \left[f''(\phi_{1j}) + \frac{f'(\phi_{1j})}{\phi_{1j}}\right]\right\}, \\
    \end{split}
    \label{eq:gamma_min}
\end{equation}
As found out in Ref. \cite{asenjo2017exponential}, the contribution of the point-like atoms (first term) decreases exponentially with the number of atoms. The contribution from the spread of the atoms is given by
\begin{equation}
    \begin{split}
        \gamma_{\text{spread}} &= 2 \eta^2 \sum_{j=2}^{N} (-1)^j \text{Im}\left[f''(\phi_{1j}) + \frac{f'(\phi_{1j})}{\phi_{1j}}\right] \\
             &\overset{N \rightarrow \infty} \approx {C \gamma_{0} \eta^2},
    \end{split}
    \label{eq:gamma_spread}
\end{equation} which is independent of $d$ at large $N$. $C$ is a proportionality constant that depends on the polarization. For a transition dipole moment that is perpendicular to the plane of the array, $\hat{q}=\hat{z}$, $C=0.8$. For a circular polarized transition, $\hat{q}=(\hat{x}+\text{i}\hat{y})/\sqrt{2}$, $C = 0.6$. For other polarization directions in the $x-y$ plane, it was numerically checked that $C = 0.6$, although the spin-waves are no longer the true eigenstates because the circular symmetry is violated. The constancy of the decay rate at large $N$ and its independence of $d$ is due to the form of the Green's function and its derivatives. These functions have a dominant term at large atom separation, $R_{ij}$, which is a sinc function, $\sin(\phi_{ij})/\phi_{ij}$. The alternating sum in Eq. (\ref{eq:gamma_spread}) of this function has a flat behavior at large $N$ for all $d < 0.5\lambda_0$.

The above analysis assumes no $z$ spread. However in realistic scenarios, the atoms have a finite spread in the $z$ direction, which can also happen to be different from the spread in the $x-y$ plane. This $z$ spread has a negligible effect on the forces, but affects the limiting decay rate. We denote the LD parameter in the $z$ direction as $\eta_z$, while the LD parameter in the $x-y$ plane is $\eta \equiv \eta_{xy}$. The limiting decay becomes relatively complicated. To illustrate, for $\hat{q}=\hat{z}$ the limiting decay rate can be expanded as
\begin{equation}
    \begin{split}
        \gamma_{\text{spread}} &= 2 \sum_{j=2}^{\infty} (-1)^j \left[\eta_{xy}^2 m(\phi_{1j}) + \eta_z^2 n(\phi_{1j})\right],
    \end{split}
    \label{eq:gamma_spread_z}
\end{equation} where if we denote $c \equiv cos(x), s \equiv sin(x)$ the functions \( m \) and \( n \) are given by
\begin{equation}
    \begin{split}
        m(x) &= \frac{- 1.5 x^{4} s - 3.0 x^{3} c + 7.5 x^{2} s + 13.5 x c - 13.5 s}{x^{5}}
        , \\
        n(x) &= \frac{1.5 x^{3} c - 6.0 x^{2} s - 13.5 x c + 13.5 s}{x^{5}}
        .
    \end{split}
    \label{eq:mn} 
\end{equation} In this case, the function $m$ contributes a factor of $0.8$, while the function $n$ contributes a factor of $0.2$ to the decay rate. Thus, for isotropic spread, the proportionality constant $C$ becomes $1$, while for $\eta_z = 2\eta_{xy}$, $C = 1.6$.

\bibliography{main}

\end{document}